\documentclass{ptapap}
\usepackage[]{natbib}

\author{Joyce A. Guzik}[LANL]
\author{Chris Fryer}[LANL]
\author{Todd J. Urbatsch}[LANL]

\author{Stanley P. Owocki}[UDel]
\affil[LANL]{Los Alamos National Laboratory,  Los Alamos, NM 87545, USA}
 \affil[UDel]{Dept. of Physics and Astronomy, Bartol Research Institute, University of Delaware, Newark, DE 19716, USA}

\title{Radiation Transport Through Super-Eddington Stellar Winds}
\headtitle{Radiation Transport Through Super-Eddington Stellar Winds}

\begin{document}

\maketitle

\begin{abstract}

We present results of simulations to assess the feasibility of modeling outflows from massive stars using the Los Alamos 3-D radiation hydrodynamics code Cassio developed for inertial confinement fusion (ICF) applications.  We find that a 1-D stellar envelope simulation relaxes into hydrostatic equilibrium using computing resources that would make the simulation tractable in 2-D.  We summarize next steps to include more physics fidelity and model the response to a large and abrupt energy deposition at the base of the envelope.

\end{abstract}

\section{Introduction}

Our goal is to explore the consequences for eruptive mass loss in massive stars that may be driven by an abrupt high energy-deposition rate in the interior.  This energy deposition could be initiated by, e.g., gravity waves or binary merger and common envelope evolution.  These mechanisms have been proposed to explain the enhanced mass loss during giant eruptions of luminous blue variables (LBVs) and in pre-supernova events.

The radiation flow and hydrodynamics may cause the outer layers to break up into clumps or become `porous', reducing the effect of radiation driving and changing stellar wind properties and other observational characteristics.   3-D hydrodynamic models with radiation transport methods that are more sophisticated than radiation diffusion will be needed to model the outcome.

We present modeling of this problem using the Los Alamos code Cassio, an Eulerian adaptive-grid radiation hydrodynamics code that includes implicit Monte Carlo (IMC) radiation transport.  This code is capable in principle of modeling both radiative and convective energy transport, and accounting for radiation flow through inhomogeneous clumps and density gradients created from outflow and subsequent fallback.  We would like to compare results with the analytical work of \cite{2017MNRAS.472.3749O} and prior 1-D and 3-D simulations.

\section{Cassio simulations}

We modeled the problem in 1-D planar geometry, with minimum zone size 8.75$\times10^6$ cm.  The box size of the simulation was 7$\times10^{10}$ cm ($\sim$1 R$_{\odot}$).  We used the two-temperature (material and radiation temperature) option, and diffusive radiation transport.  The code temperature units are electron volt (eV); 1 eV is equivalent to 11,604 K.  We used a 50\% H/50\% C mixture for the Rosseland mean opacity and equation of state, carried over from an ICF calculation of CH foam.  We introduced a downward constant gravitational acceleration of 110 cm s$^{-2}$, equivalent to that of a 30 M$_{\odot}$ star at distance 86.5 R$_{\odot}$.  The radius and mass of the star were estimated from Z=0.004 mass-losing models of initial mass 50 M$_{\odot}$ \cite[see, e.g.,][]{1993A&AS..101..415C} that are evolved to near the LBV instability region.  We initiated the matter with a uniform material density 0.02 g cm$^{-3}$ of thickness 7$\times10^9$ cm ($\sim$1/10 R$_{\odot}$).  The remainder of the mesh was filled with air at density at 10$^{-8}$ g cm$^{-3}$, again carried over from the ICF problem setup.  We initiated the material at 4.3 eV ($\sim$50,000 K), typical of the photosphere of an evolved massive star, and also introduced a bottom constant-temperature boundary condition of 4.3 eV.  The outer boundary has a vacuum boundary condition, allowing outflow of mass and radiation.

The code uses several criteria to limit the time step; a typical time step near the beginning of the simulation is $\sim$1 s.  As the problem relaxes into equilibrium, the time step increases gradually, and reaches $\sim$3.5 s after 70 days of stellar time.  The time step is limited by a velocity constraint on movement of material across a cell boundary to $<$20\% of the cell size.  The run to 70 days stellar time required wall-clock time 42 hours on two nodes (16 cores per node).  The typical number of cells is $\sim$1250.

\section{Model expectations and results}

We expect that the gravitational field should cause the matter to develop a density stratification in which the pressure gradient balances the downward gravitational force (hydrostatic equilibrium).  Also, the bottom boundary condition should generate a constant luminosity, i.e., a constant energy transport across the mesh, with a constant radiation energy outflow from the upper boundary of the problem (thermal equilibrium).  

We calculated the dynamical and thermal timescales for the chosen initial conditions in order to determine roughly how long these relaxations should require.

For
\begin{equation}
g = (GM)/R^2 = 110 {\rm~cm~s}^{-2},
\end{equation}

\noindent the dynamical timescale of the envelope is
\begin{equation}
 t_{dyn} = (R^3/GM)^{0.5} = ({\Delta}R/g)^{0.5} = ((7\times10^9)/110)^{0.5} = 7.977\times10^3~{\rm s} \simeq 2.2~{\rm hr}. 
\end{equation}

\noindent The thermal (Kelvin-Helmholtz) timescale is
\begin{equation}
t_{K-H} = GM{^2}/RL = g{\Delta}R{\Delta}M/L = 110(7\times10^9)(1.54\times10^8)/1.36\times10^{12}
\end{equation}
\indent  $= 8.7\times10^7~{\rm s} \simeq$ 1009 days.

\begin{figure}
\center
    \includegraphics[width=0.35\textwidth]{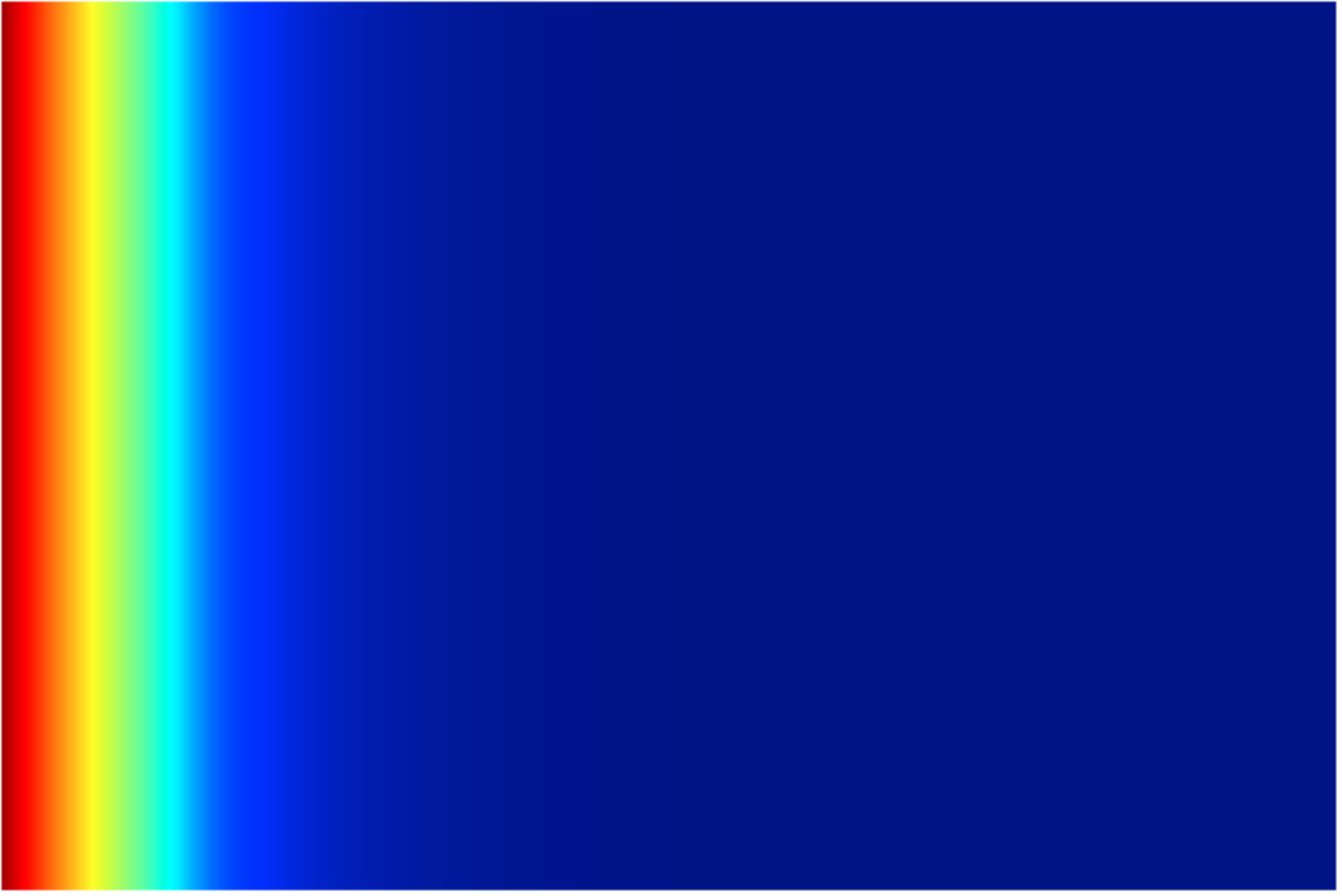}
     \includegraphics[width=0.35\textwidth]{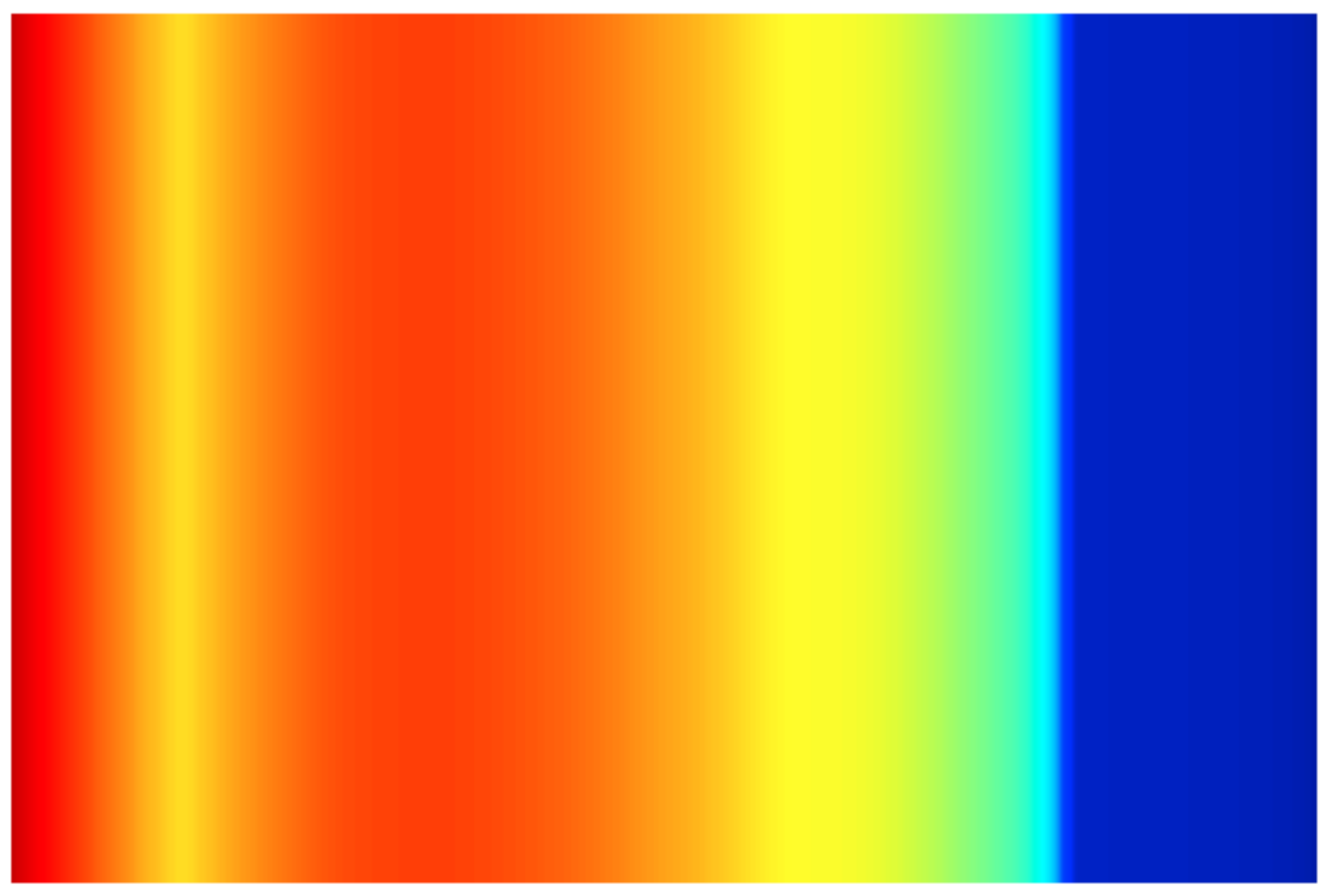}
    \includegraphics[width=0.38\textwidth]{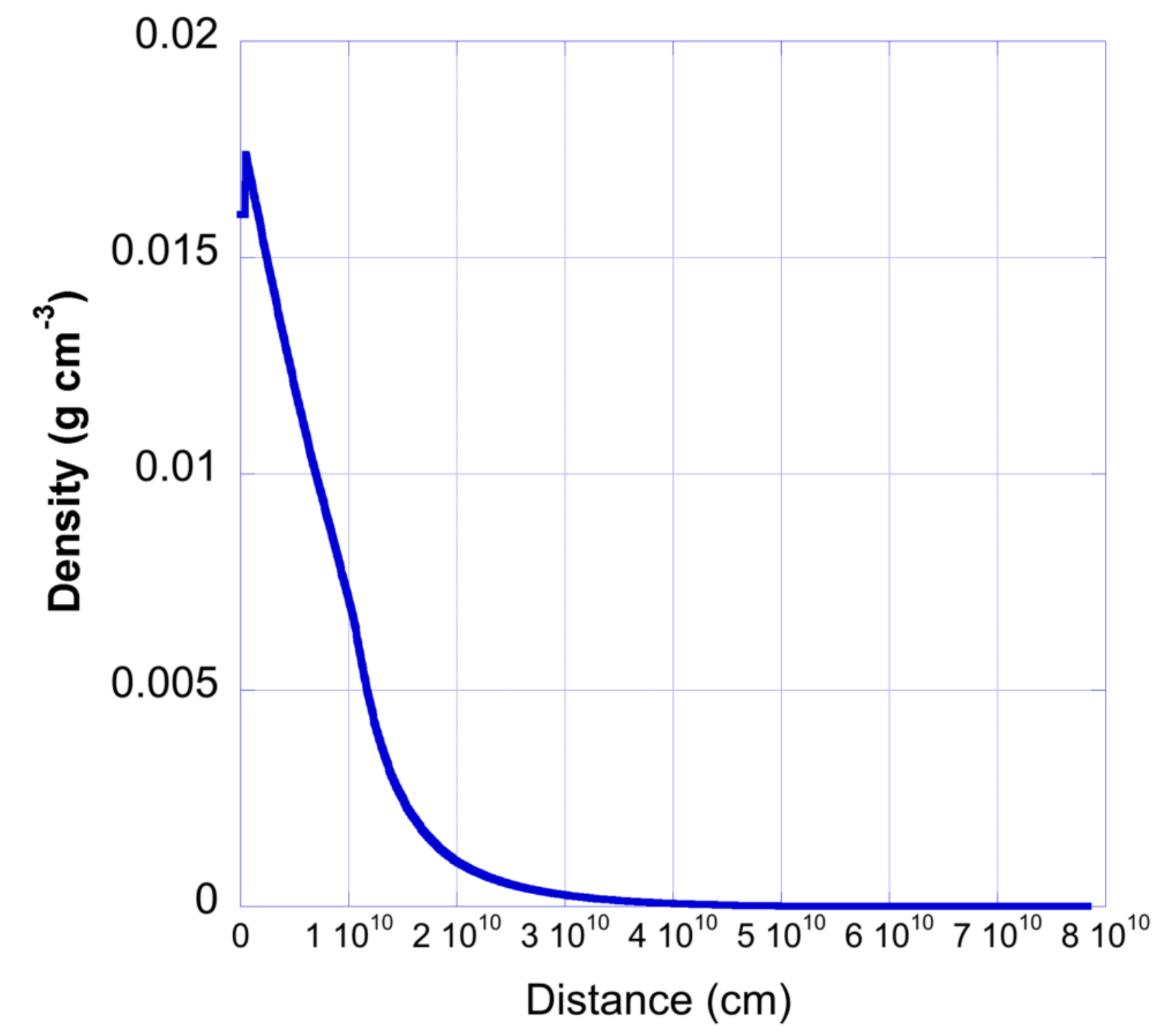}
    \includegraphics[width=0.35\textwidth]{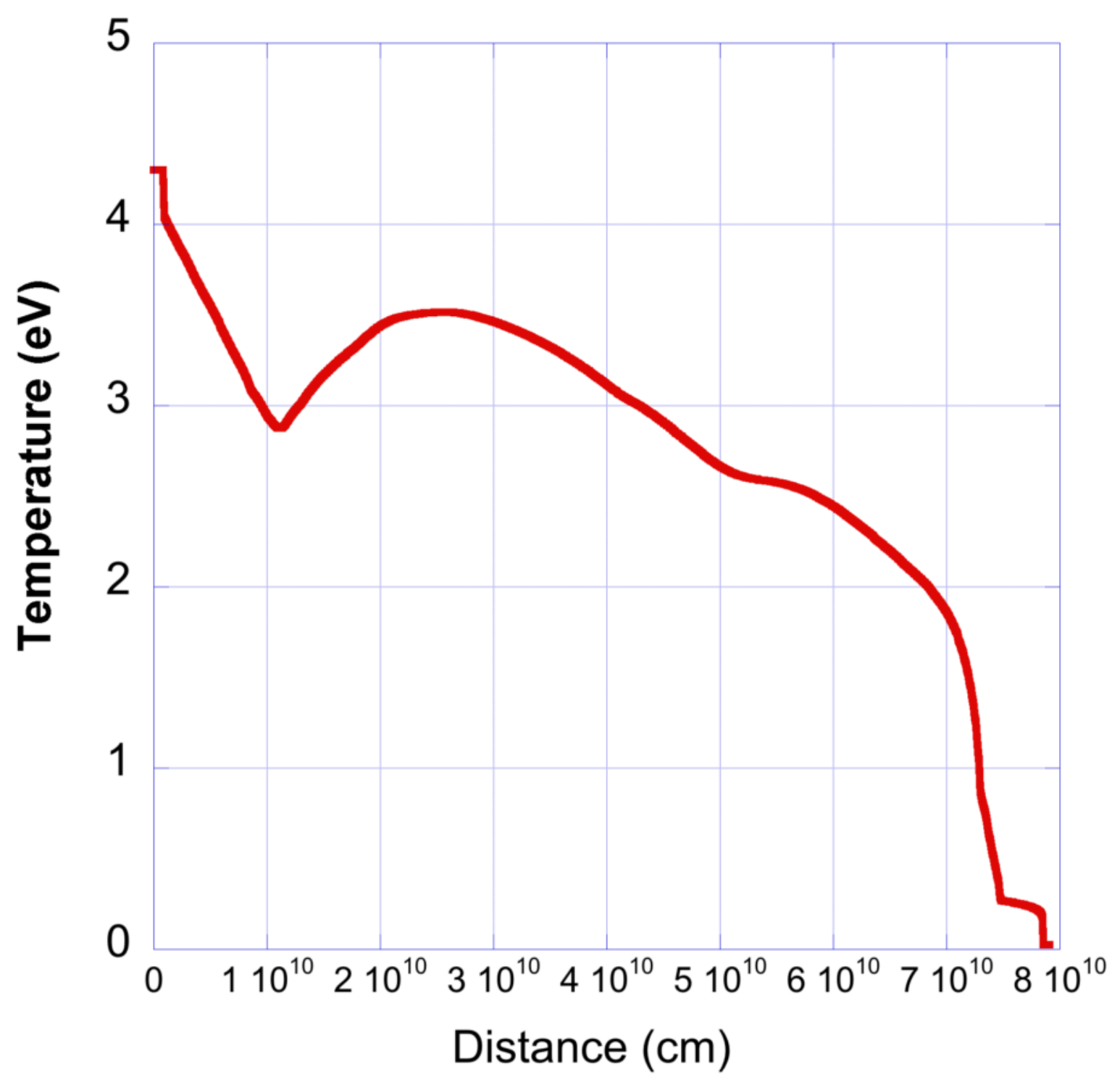}
    \caption{Density and temperature contours (top) and profiles (bottom) for simulation at 70 days. The density profile has settled into a nearly hydrostatic configuration.  The temperature profile has features remaining from a heat wave moving outward (to the right).}
    \label{fig:Cassio}
\end{figure}

\vspace{2mm}
Figure \ref{fig:Cassio} shows density and temperature contour plots and profiles at 70 days.  The density gradient is nearly smooth.  However, the temperature gradient has some structure that needs to be investigated, and a heat front is still moving outward (to the right), possibly because of very slow thermal conduction in this low-density region.  There is also a temperature inversion at $\sim$10$^{10}$ cm that may be caused by some feature of the opacity or equation-of-state treatment, or possibly is the consequence of extrapolating off of a table boundary. 

For this simulation, 1.3$\times10^6$ g out of the initial mass of 1.54$\times10^8$ g ($<$1\%) leaves the problem, driven by shocks generated from the initial mass collapse.  The rate that energy leaves the problem becomes nearly constant at 1.36$\times10^{12}$ erg cm$^{-2}$ s$^{-1}$.  Because this is a planar problem, approximating a portion of stellar envelope at 86.5 R$_{\odot}$ from the stellar center, this rate of energy loss translates to about 1.6$\times10^5$ L$_{\odot}$.

\section{Conclusions and next steps}

The LANL Cassio code shows promise for being able to run this problem, and generate reasonable initial conditions.  We were able to show that the 1-D simulation will relax into a steady state in hydrostatic and thermal equilibrium in a timescale that may be possible to reach in the simulations.

Some next steps are to zone the problem in a 2-D, possibly spherical, geometry.  After the simulation settles into equilibrium, we would introduce a high energy-deposition rate at the base of the stellar envelope to generate super-Eddington luminosity, and investigate whether convection, clumpy winds, or porosity develop.  We would like to compare results using IMC radiation transport vs.~radiation diffusion.  We could replace the 2-T model with a 3-T (radiation, ion, electron temperature) model.  We would use a more realistic composition, e.g., H/He + trace amounts of CNOFe elements, as appropriate for stellar envelopes.  We could consider developing links from stellar evolution codes to set the initial conditions.

\acknowledgements{We thank the KITP Stars 2017 program for support that facilitated this collaboration, and the Department of Energy's High Energy Density Physics Impact program for support to attend this conference.}

\bibliographystyle{ptapap}          
\bibliography{Guzik2}

\end{document}